# Temperature-Driven Structural Phase Transitions in $SmNiO_3$: Insights from Molecular Dynamics Simulations

Guoyong Shi,[1] Fenglin Deng,[2,3] Ri He,[4,5] Dachuan Chen,[6] Xuejiao Chen,[7] Sen Yang,[1] Zhicheng Zhong[2,3], and Peiheng Jiang[1,*]

[1] *School of Physics, MOE Key Laboratory for Nonequilibrium Synthesis and Modulation of Condensed Matter, Xi'an Jiaotong University, Xi'an 710049, China*

[2] *School of Artificial Intelligence and Data Science, University of Science and Technology of China, Hefei 230026, China*

[3] *Suzhou Institute for Advanced Research, University of Science and Technology of China, Suzhou 215123, China*

[4] *Ningbo Institute of Materials Technology and Engineering, Chinese Academy of Sciences, Ningbo 315201, China*

[5] *College of Materials Science and Opto-Electronic Technology, University of Chinese Academy of Sciences, Beijing 100049, China*

[6] *Ningbo Institute of Digital Twin, Eastern Institute of Technology, Ningbo 315201, China*

[7] *School of Photoelectric Engineering, Changzhou Institute of Technology, Changzhou, Jiangsu, 213002, China*

## ABSTRACT

The metal-insulator transition (MIT) in rare-earth nickelates exemplifies the intricate coupling between lattice dynamics and electronic effects. This strong interplay makes it challenging to disentangle their individual roles in driving the transition in $R$NiO$_3$. Here, we isolate the structure response from electronic effect by employing molecular dynamics (MD) simulations based on a machine-learned interatomic potential. Taking $SmNiO_3$ as a prototypical system, our simulations show that the structural phase transition is intrinsically temperature-driven and occurs spontaneously via collective lattice distortions. The simulated critical temperature is 340 K and can be further tuned by pressure. These findings provide atomistic insights into the understanding of structural evolution in triggering the phase transition and hence the MIT in $R$NiO$_3$.

[*] jiangph@xjtu.edu.cn

# I. INTRODUCTION

Rare earth nickelates ($R$NiO$_3$ with $R$ = rare-earth, except La), as prototypical strongly correlated oxides, exhibit a sharp temperature-driven metal-insulator transition (MIT) [1,2]. This transition is accompanied by both structural and electronic transition, i.e., bond disproportionation and charge disproportionation [3-6]. The bond disproportionation depicts the breathing-mode distortion of the NiO$_6$ octahedra, which concurrent with a symmetry reduction from the high temperature orthorhombic *Pbnm* symmetry to the low temperature monoclinic $P2_1/n$ symmetry [7-10]. The charge disproportionation, occurred simultaneously, is illustrated by $2(d^8L^1) \rightarrow d^8 + d^8L^2$ ($L$ denotes a ligand hole of oxygen) [11-14], corresponding to the splitting of Ni$^{3+}$ cations to Ni$^{2+}$ and Ni$^{4+}$ states. These two disproportionation effects, which are strongly coupled, are believed to play a key role in the onset of the phase transition and hence the MIT [3,15,16]. During the transition, the emergence of bond and charge disproportionation weakens the *d-p* hybridization and reduces the bandwidth, leading to the opening of a band gap [3,17].

The complex ordering induced effect in $R$NiO$_3$ has attracted significant interest in the past three decades. Among all investigations, one of the key questions is the driving force of the phase transition. Early studies attributed the phase transition in $R$NiO$_3$ primarily to correlated electronic effects [11,12,18,19]. In this picture, charge instability—such as interorbital charge fluctuations—drives the system into a charge disproportionate state, and then the bond disproportionation is induced simultaneously through strong electron-lattice coupling [20]. However, recent density functional theory (DFT) calculations underscore the crucial role of the lattice structure [21-23]. These calculations indicate that the bond disproportionation is triggered by the softening of an oxygen-breathing distortion, and then result in the emergence of charge disproportionate state. Therefore, the intricate interplay between structure and electronic behavior, whether cooperating [24] or competing [25], remains elusive. Consequently, the physical mechanism driving the phase transition in $R\mathrm{NiO_3}$ remains debated.

It is subtle and challenging to disentangle the structural and electronic effect in the phase transition of $R\mathrm{NiO_3}$. In the DFT+$U$ scheme, both the structure and electronic behavior can be completely described, including the metallic phase in *Pbnm* symmetry and insulating phase in $P2_1/n$ symmetry with the bond and charge disproportionation [26]. The structure phase transition can be even characterized by *ab-initio* molecular dynamic (AIMD) simulations [27,28]. However, the structural and electronic states are treated in an unbiased way in the DFT+$U$ scheme, so that their responses are inherently entangled, making it difficult to decouple their individual roles in the phase transition.

To address this challenge, we employed machine-learned interatomic potentials with multiscale molecular dynamics (MD) to investigate the phase transition of $\mathrm{SmNiO_3}$−a representative nickelate that displays a pronounced phase transition near room temperature. The interatomic potentials can be obtained by training neural networks on DFT+$U$ datasets. With this approach, we can decouple the structure response from the complete description (both structure and electronic) with the same accuracy. Our MD simulations clearly display the temperature-driven structural evolution of $SmNiO_3$. The bond disproportionation effect, induced by breathing-mode distortion, can be well reproduced in our approach. The disproportionation is gradually suppressed with increasing temperature and disappears at the critical temperature of 340 K. This critical temperature can be further modulated by pressure. Our work provides a new insight for understanding the physical mechanisms of the phase transition and hence the MIT in $RNiO_3$.

## II. METHOD

Our DFT calculations were performed with the Vienna *Ab initio* Simulation Package (VASP) by adopting the projector augmented wave (PAW) method [29-31]. The exchange correlation functional was treated using the generalized gradient approximation (GGA), as formulated by PBEsol [32]. The valence configurations were employed as following: $5s^2 6s^2 5p^6 5d^1$ for Sm, $3d^9 4s^1$ for Ni, and $2s^2 2p^4$ for O. The DFT+$U$ method with a Hubbard correction ($U_{eff} = U - J = 2.0$ eV) was applied to describe the $3d$ orbitals of Ni [4,26,27,33-35]. Our $U$ value test calculations indicate that this moderate $U$ value accurately describes the structural and electronic behaviors of $SmNiO_3$. The energy cutoff of 500 eV was adopted for the plane-wave basis set. The Monkhorst-Pack $k$-point grid was optimized to sample the Brillouin zone with a grid spacing of 0.16 Å$^{-1}$ $k$-points [36]. The spin-polarization was considered in both ground state calculations and AIMD simulations.

The general Deep Potential Generator (DP-GEN) scheme [37] and the DeepPot-SE [38] were utilized to generate the training datasets. We first constructed supercells consisting of 40 atoms corresponding to the $P2_1/n$ and $Pbnm$ space group, and the T-type antiferromagnetic (AFM) and ferromagnetic (FM) configurations are adopted for these two structures, respectively. Subsequently, each supercell was randomly perturbed by scaling the lattice translation vectors with a range of -3 to 3% and by applying relative atomic displacements within a range of -0.008 to 0.008 Å. Next, 5-step AIMD simulations were performed for each perturbed structure in the *NVT* ensemble at 50 K, and a total of 400 ionic configurations were collected. During each DP-GEN loop training step, Deep potential (DP) model was trained using randomly initialized neural network parameters. Each training step comprised 600,000 steps. A cutoff radius of 6.0 Å and a smooth cutoff radius of 2.0 Å were employed. In the exploration step of the DP-GEN loop, DP model is selected to investigate various structures of $SmNiO_3$ in $P2_1/n$ and *Pbnm* supercells using Deep Potential Molecular Dynamics (DPMD) with the LAMMPS package [39]. MD simulations were conducted within the isothermal-isobaric (*NPT*) ensemble over a temperature range of 50 − 600 K. A time step of 1 fs and periodic boundary conditions in all directions were adopted in our MD simulations. Finally, we trained a unified potential model for describing both $P2_1/n$ and *Pbnm* phases. For detailed information on each step, please refer to our previous works [40,41].

## III. RESULTS AND DISCUSSION

We first investigate the relationship between the structure and electronic properties of $SmNiO_3$ in the DFT+*U* scheme. The calculations were performed on 80-atom supercells using T-type antiferromagnetic (AFM) and ferromagnetic (FM) configurations for the $P2_1/n$ and *Pbnm* phases, respectively [4]. As shown in Fig. 1(a) and (b), the fully relaxed lattice parameters are $a$ = 5.386 Å, $b$ = 10.511 Å, $c$ = 14.966

Å for $P2_1/n$ phase and $a$ = 5.407 Å, $b$ = 10.495 Å, $c$ = 14.952 Å for *Pbnm* phase, respectively. The $P2_1/n$ phase exhibits a breathing-mode distortion of the $NiO_6$ octahedra, and hence includes two distinct Ni states, which are classified as expanded (large, donated as "L") and compressed (small, donated as "S") units as shown in Fig. 1. The fully relaxed volumes of L and S octahedral units are 10.08 $Å^3$ and 9.06 $Å^3$, respectively. In contrast, the *Pbnm* phase exhibits equivalent Ni state with uniformly $NiO_6$ octahedra (middle, denoted as "M"), with each octahedron having a volume of 9.60 $Å^3$. Total energy calculations show that the energy of $P2_1/n$ phase is 2.99 meV/atom lower than that of *Pbnm* phase, which indicating that $P2_1/n$ phase is the ground state at low temperature. Figs. 1(c-d) plot the calculated density of states (DOS) for both phases. The $P2_1/n$ phase is an insulator with a band gap of 0.474 eV, whereas the *Pbnm* phase displays metallic characteristics. This result agrees very well with previous theoretical and experimental works [4,26,42,43], and in accordance with the phase transition scenario that metallic *Pbnm* state transition to insulating $P2_1/n$ state at critical temperature. We note that although DFT+$U$ does not fully capture the strongly correlated physics in nickelates [44,45], it still reproduces the essential structural and electronic features of $SmNiO_3$ [3,26], and is therefore sufficient for generating the training datasets for interatomic potentials.

To further reveal the linkage between electronic behavior and lattice structure, we examine the DOS of various structures with different symmetries and distortion patterns, and confirm that the band gap emerges simultaneously when the bond disproportionate state is included. This result reveals that the structure feature of bond disproportion is strongly coupled with the opening of band gap in $R$NiO$_3$ [3,26,46,47]. Therefore, understanding the evolution of bond disproportionation is crucial for unraveling the driving force of the phase transition.

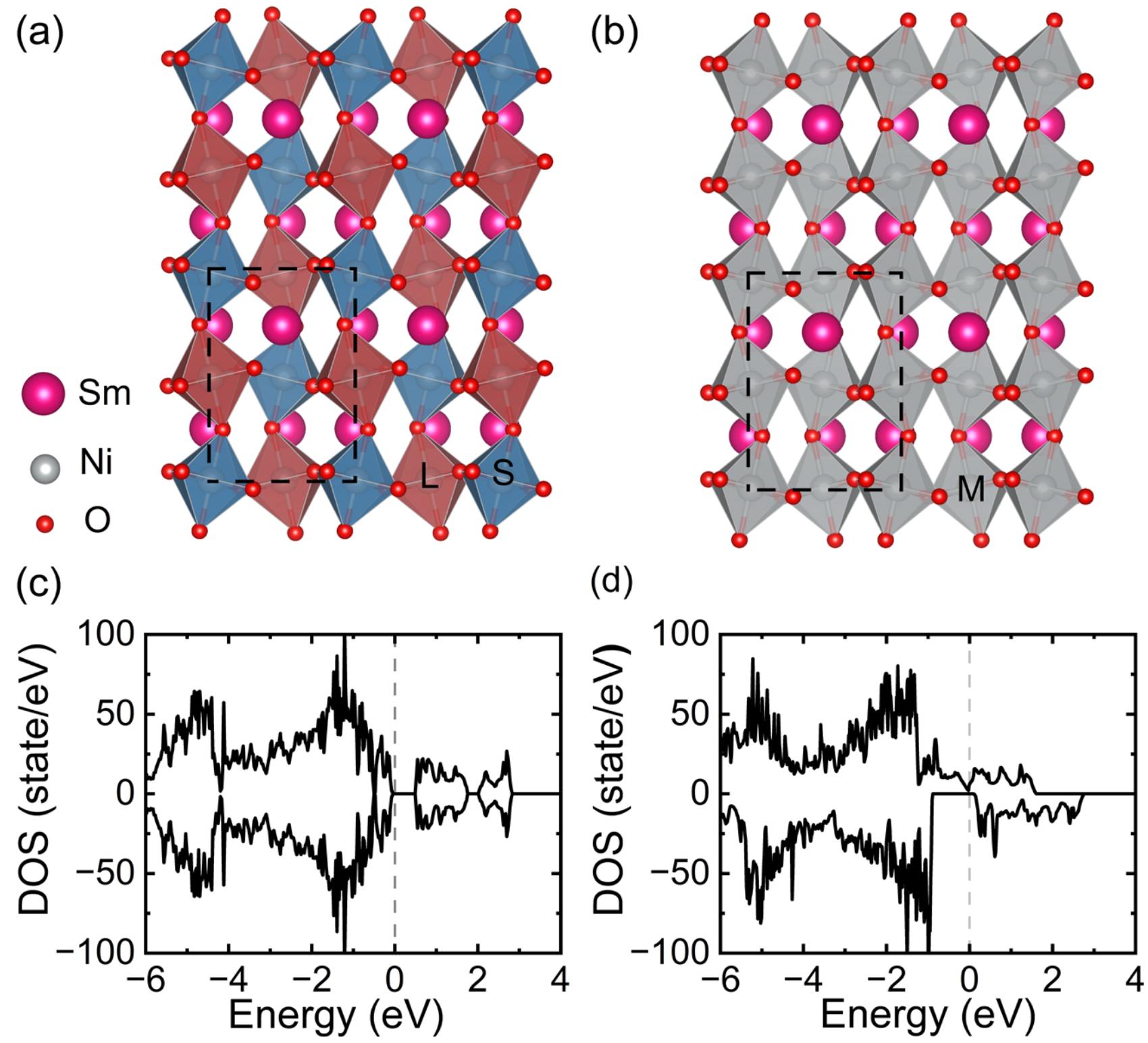

Fig. 1. Crystal structures of SmNiO$_3$ in (a) $P2_1/n$ phase and (b) $Pbnm$ phase. Sm, Ni, and O atoms are represented by pink, grey, and red spheres, respectively. According to the volume, the NiO$_6$ octahedra are categorized as large ("L") and small ("S") for $P2_1/n$ phase and middle ("M") for $Pbnm$ phase, and are colored by red, blue and grey, respectively. The corresponding density of states (DOS) for $P2_1/n$ and $Pbnm$ phases are shown in panels (c) and (d), the Fermi energy is set to 0 eV.

Based on the DFT+$U$ results, we next investigate the evolution of bond disproportion in SmNiO$_3$ with DPMD approach. The DP model was constructed within a supervised machine learning framework. The training dataset comprises structural information, total energies, atomic forces, and virial tensors obtained from spin-polarized DFT+$U$ calculations. The final dataset, generated via the DP-GEN iterative procedure, contains 1429 configurations. A comparison of the energies and forces predicted by the DP model with those calculated via DFT+*U* are illustrated in Fig. 2. The root-mean-square errors (RMSE) for energy and force are 2.03 meV/atom and 84.74 meV/Å (*x*-axis), 80.99 meV/Å (*y*-axis), and 84.62 meV/Å (*z*-axis), respectively. Such RMSEs are smaller than the general criterial of ~10 meV/atom and ~ 100 meV/Å, respectively [48], and hence validate the accuracy of the interatomic potentials in DP model and the robustness of the training protocol.

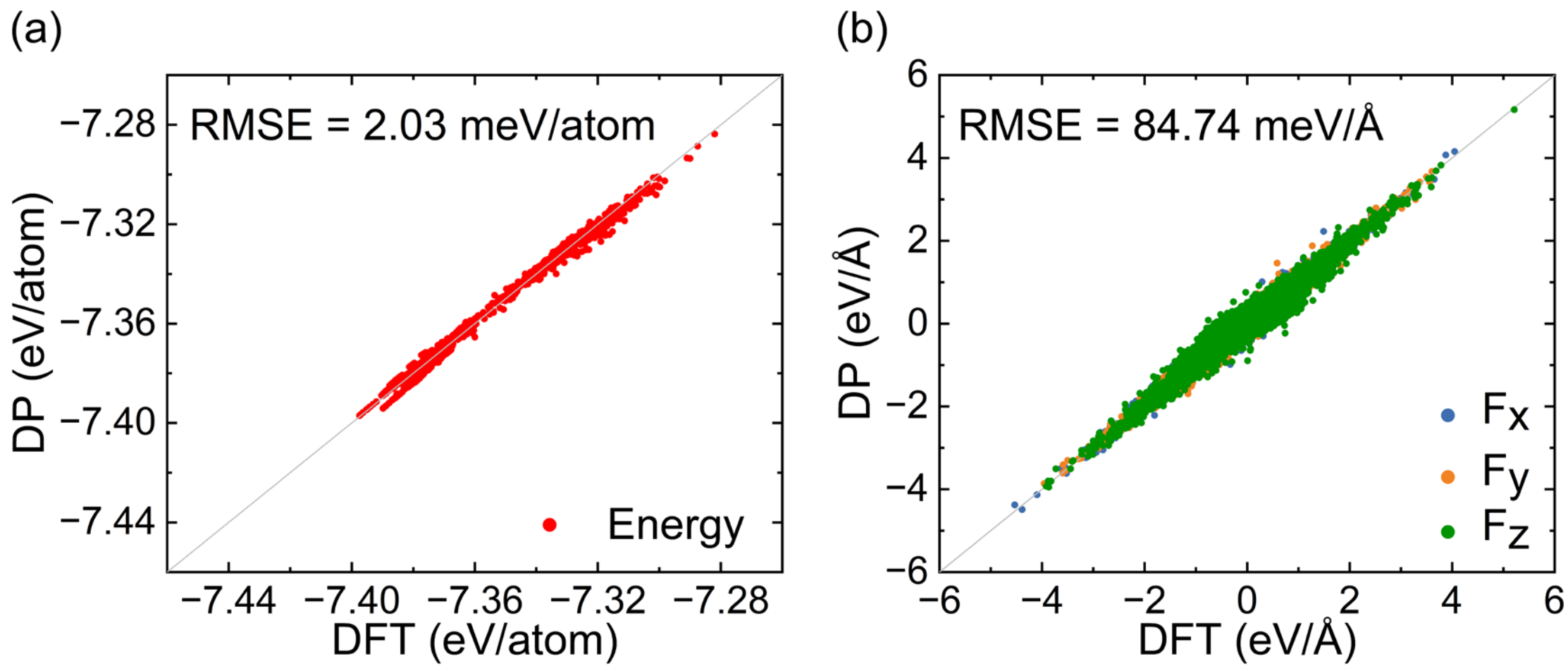

Fig. 2. Comparison between DFT+*U* calculations and DP model for $SmNiO_3$: (a) total energies and (b) atomic forces.

After validating the potential model, MD simulations were performed on $SmNiO_3$ using a 7680-atom supercell. The simulated bond disproportion evolution is shown in Fig. 3. At low temperature of 50 K, the significant difference in volume between the small and large $NiO_6$ octahedra indicates the persistence of the breathing-mode distortion characteristic, and agrees with the bond disproportion feature of the low-temperature $P2_1/n$ phase. At elevated temperature, the enhanced thermal fluctuation broadens the volume distribution of $NiO_6$ octahedra, leading to a partial overlap between small and large octahedra at 150 K. Notably, at the higher temperature of 320

K, the peak value separation in volume between small and large octahedra is substantially reduced, suggesting a strong suppression of bond disproportion. Upon further heating to 340 K, these two volume distribution curves merge completely, indicating that the previous two kinds of octahedra become to the equivalent one, and hence bond disproportion disappears, which leading to the structure transitions to the centrosymmetric *Pbnm* phase. In addition, we also simulated the opposite process, i.e., the cooling process, with the start point of high temperature *Pbnm* phase. When the temperature decreases to 340 K, the bond disproportion emerges. This disproportion effect is strengthened at low temperature and the structure transitions to $P2_1/n$ phase finally. All these simulations indicate the robustness of the temperature driven phase transition and bond disproportion evolution in $SmNiO_3$.

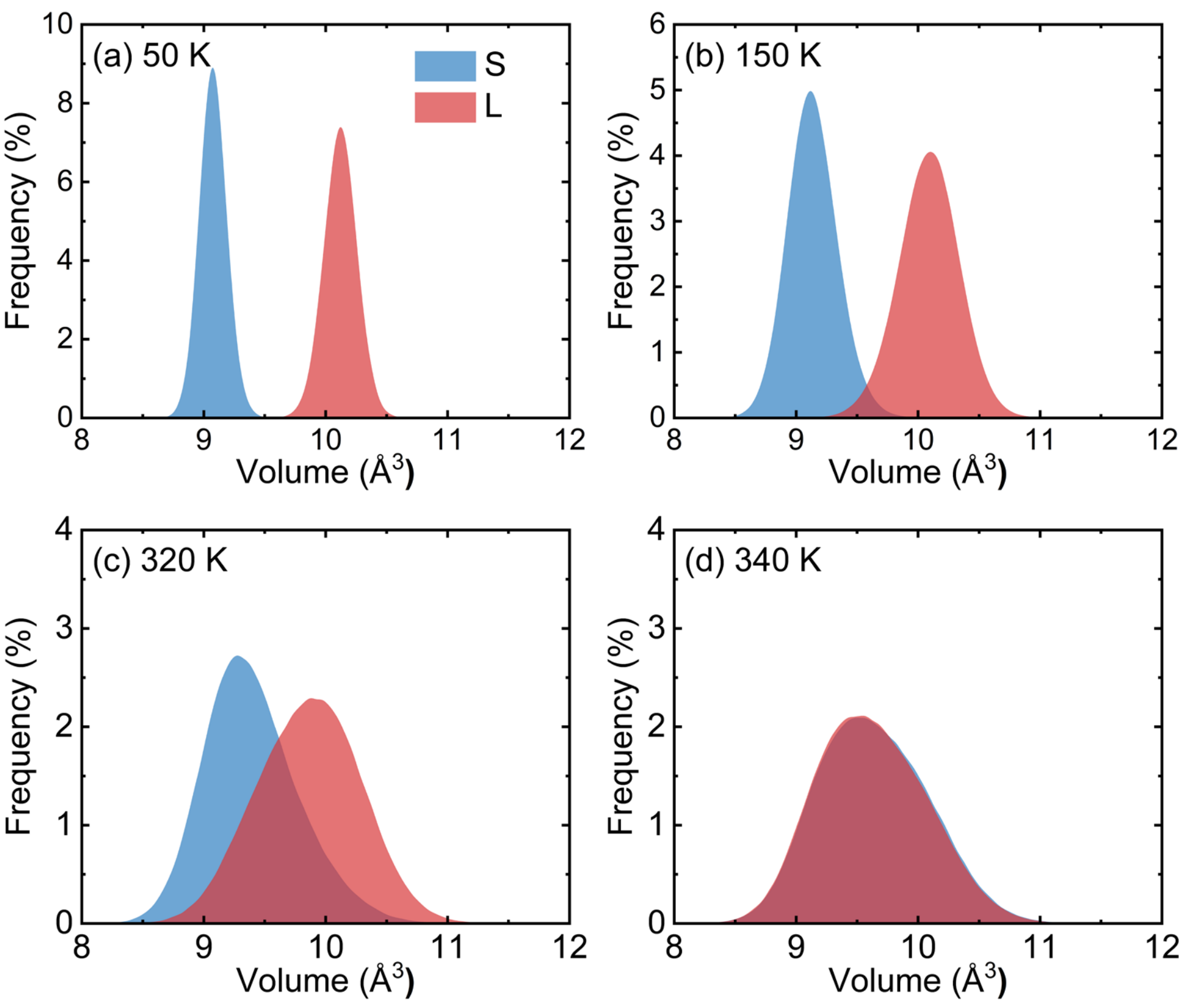

Fig. 3. The MD simulated distribution profile of $NiO_6$ octahedral volume in $SmNiO_3$. The S and L represent the volume distribution of small and large $NiO_6$ octahedra, respectively.

To further demonstrate the temperature-dependent structural evolution, we conducted crystallographic analysis by extracting the final simulated atomic configurations at different temperature and projecting them along the [001] crystallographic axis. Fig. 4 shows that at 5 K, the $NiO_6$ octahedra form a well-ordered, three-dimensional checkerboard pattern. This arrangement, characterized by a breathing-mode distortion in the $P2_1/n$ phase, maintains long-range order via the periodic modulation of octahedral sizes at low temperatures. As the temperature increases, the breathing-mode distortion is progressively weakened, and is eventually vanished at high temperature (e.g., at 500 K). The loss of such characteristic means that the expanded and contracted octahedra becomes indistinguishable and all $NiO_6$ units converging toward a uniform configuration in the *Pbnm* phase. We note that at high temperature, the random distribution of octahedra volume increases, which should be attributed to the significantly increased thermal fluctuations and dynamic process.

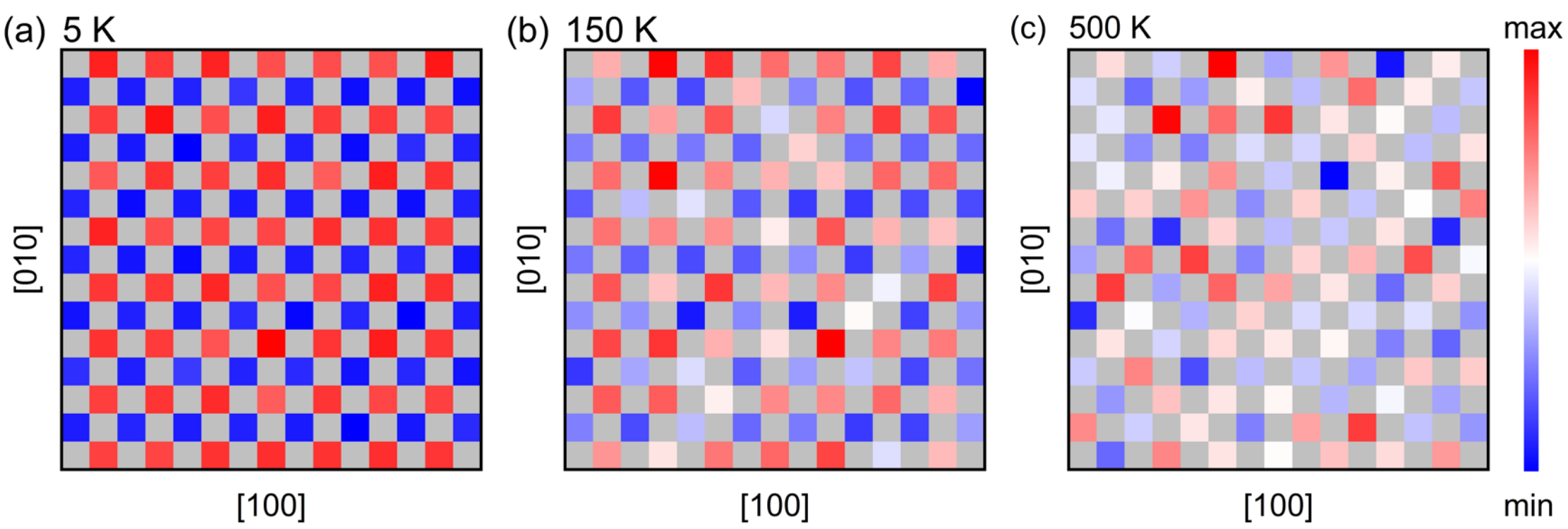

Fig. 4. The $NiO_6$ octahedral volume mapping along the [001] crystallographic direction at (a) T = 5 K, (b) T = 150 K, and (c) T = 500 K. Each marker corresponds to the real-space coordinates of a $NiO_6$ octahedron center, with data points color-mapped according to their octahedral volume. The right bar is color scale of volume.

To determine the critical temperature of the phase transition, we extract the peak value from each octahedral volume distribution curve in Fig. 3 and plot them as a function of temperature in Fig. 5(a). At low temperatures, significant octahedral disproportionation is observed, characterized by a marked volume difference between the large and small units — a signature of the breathing-mode distorted $P2_1/n$ phase. As the temperature increases, this structural disproportionation is progressively suppressed, and the volume of two kinds of octahedral converges to the same value at 340 K. This temperature is identified as the critical temperature for the transition from the $P2_1/n$ phase to the centrosymmetric *Pbnm* phase. It should be noted that classical MD simulations do not explicitly include electronic degrees of freedom. This approach differs from previous DFT+*U* and AIMD simulations, where the electronic state is spontaneously coupled with lattice structure. Therefore, the phase transition observed in our simulations is primarily driven by structural variations which are amplified by thermal fluctuations. On the other hand, compared with experimental observations, our simulated critical temperature is lower than the experimental value (~ 400 K), and the phase transition is not as sharp as expected for this first-order transition, the hysteresis is also not observed. These discrepancies should be attributed to the absence of electronic contribution in our simulations. Our work thus indicates the cooperating behavior of lattice and electronic mechanism in describing the nature of the phase transition and hence the MIT in nickelates. This finding not only aligns with previous work [23], but also offers a clearer physical picture than that derived from recent pure DFT+*U* studies [17].

Pressure offers an alternative pathway for modulating phase transition for most of materials. Here we performed MD simulations of $SmNiO_3$ under hydrostatic pressure over a range of 0-10 GPa and illustrates the calculated pressure-temperature phase diagram in Fig. 5(b). The phase diagram shows that the critical temperature is monotonically suppressed with increasing pressure, dropping from 340 K at 0 GPa to 225 K at 10 GPa. This result agrees with previous experimental reports on $R$NiO$_3$ [49-51], thereby validating our simulations and suggesting a universal pressure-response behavior in the $R$NiO$_3$ family.

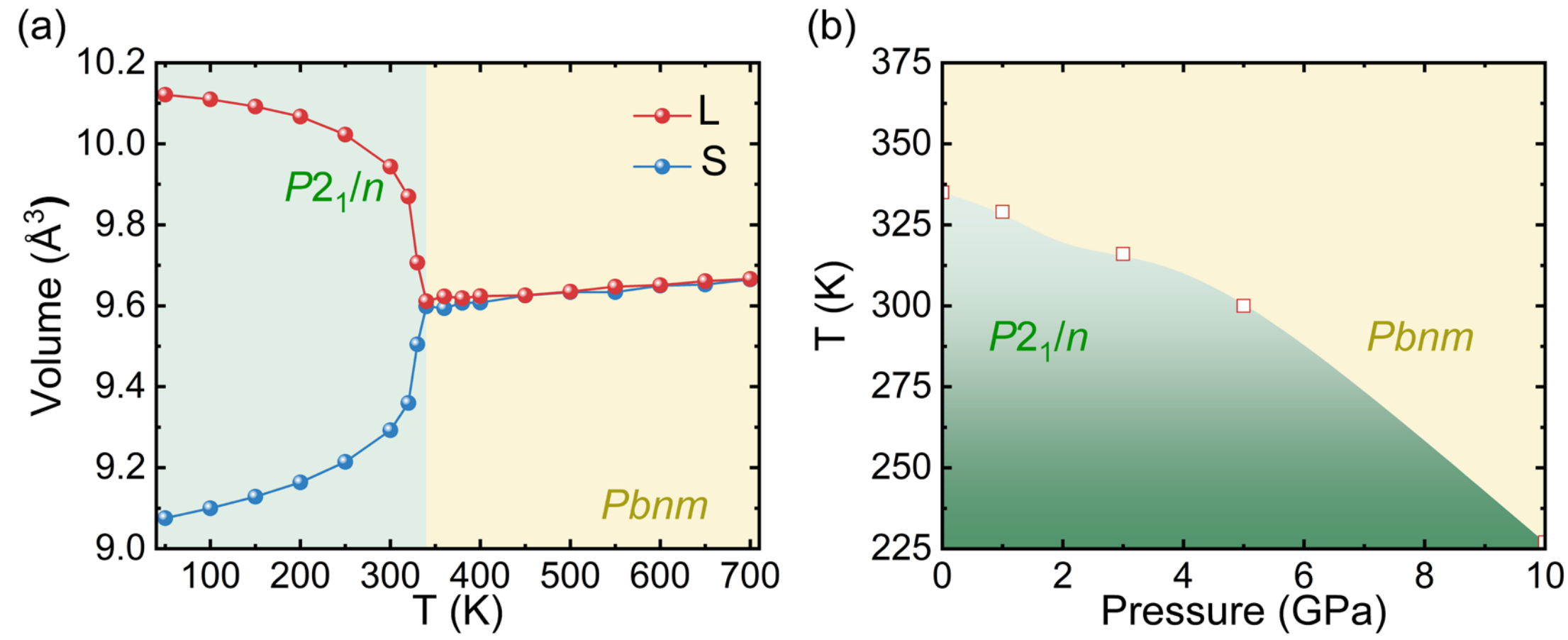

Fig. 5. (a) Temperature-dependent evolution of octahedral unit volume in SmNiO$_3$. (b) Pressure-temperature ($P$-$T$) phase diagram of SmNiO$_3$.

## IV. CONCLUSION

In summary, we have investigated the structural phase transition in SmNiO$_3$ using MD accelerated by a machine-learned interatomic potential with DFT-level accuracy. This approach enables us to isolate the structural response by decoupling it from electronic effects. We show that a classical MD framework successfully reproduces the phase transition, as demonstrated by the progressive suppression of bond disproportionation with increasing temperature and its eventual disappearance at 340 K. This finding underscores the significant role of lattice dynamics in driving the phase transition and hence the MIT of rare-earth nickelates $R$NiO$_3$, and suggests that the description of the transition could be further improved by incorporating electronic contributions.

## Acknowledgments

The authors acknowledge the financial support from the National Key R&D Program of China (Grants No. 2021YFA0718900), the National Natural Science Foundation of China (Grant No. 12574069), the Natural Science Basic Research Program of Shaanxi Province (No. 2025JCYBQN-001), Shaanxi Fundamental Science Research Project for Mathematics and Physics (No. 25JSQ008), the Young Talent Support Project of Xi'an Jiaotong University (WL6J022), the Natural Science Foundation of Zhejiang Province (LQ23A040002) and the HPC Platform at Xi'an Jiaotong University.